 \documentclass[aps,prl,twocolumn,showpacs,floatfix]{revtex4}
\usepackage{graphicx}
\def\fig#1{Fig.~\ref{fig#1}}
\begin{document}
\title{  Observation of pinning mode of stripe phases of 2D systems in high Landau levels }
\author{G. Sambandamurthy } \altaffiliation[Current address: ]{University at Buffalo, Buffalo, NY}\affiliation{ National High Magnetic Field Laboratory, Tallahassee, FL 32306, USA}
\affiliation{ Princeton University, Princeton, NJ 08544, USA}

\author{R. M. Lewis }\altaffiliation[Current address: ]{University of Maryland, College Park, MD}
\affiliation{ National High Magnetic Field Laboratory, Tallahassee, FL 32306, USA}
\affiliation{ Princeton University, Princeton, NJ 08544, USA}
\author{Han Zhu }\affiliation{ National High Magnetic Field Laboratory, Tallahassee, FL 32306, USA}
\affiliation{ Princeton University, Princeton, NJ 08544, USA}

\author{Y. P. Chen }\altaffiliation[Current address: ]{Purdue University, West Lafayette, IN}\affiliation{ National High Magnetic Field Laboratory, Tallahassee, FL 32306, USA}
\affiliation{ Princeton University, Princeton, NJ 08544, USA}

\author{L. W. Engel }
\affiliation{ National High Magnetic Field Laboratory, Tallahassee, FL 32306, USA}

\author{D. C. Tsui }\affiliation{ Princeton University, Princeton, NJ 08544, USA}
\author{L. N. Pfeiffer }
\affiliation{ Bell Laboratories, Lucent Technologies, NJ 07974, USA}
\author{K. W. West }
\affiliation{ Bell Laboratories, Lucent Technologies, NJ 07974, USA}

\date{\today}

\begin{abstract}
 We study the  radio-frequency diagonal conductivities of the anisotropic stripe phases of higher Landau levels near half integer  fillings.   In the hard direction, in which larger dc resistivity occurs, the spectrum exhibits a striking resonance, while in the orthogonal, easy direction, no resonance is discernable.   The resonance  is interpreted as a pinning mode of the stripe phase. 

\end{abstract}

\maketitle

In quantizing magnetic fields, and at low temperature, 
two-dimensional electron systems (2DES) of extremely low disorder exhibit  a striking anisotropy  in dc diagonal resistivity \cite{lillyaniso,duaniso}, 
near half integer Landau fillings $\nu =9/2,11/2,,,$ with  two or more Landau levels (LLs) completely filled.    The anisotropic states are understood as ``stripe'' states, in which  spatial charge density modulation in the form of long, thin stripes plays a role.       In addition, in ranges of $\nu$ on either side of the stripe phase,  experiments showed regions of vanishing diagonal resistance and quantized Hall resistance, like the nearest integer quantum Hall effects (IQHEs), but distinct from them.   These signatures of isotropic insulating behavior  are due to  electron solids termed ``bubble phases''  \cite{bubres}, which have been described as  triangular lattices with  clusters of $M$ carrier guiding centers at each site, but which exhibit apparent anisotropy in the presence of a dc current \cite{smetcurrent}.      For $\nu$ farther yet from half integer filling,  within the ranges of the IQHE itself,    individual carriers of the partially filled  LL  form a triangular Wigner crystal  \cite{yongiqhwc,rupertcoex} (IQHE-WC, or $M=1$)   similar to Wigner solid  states \cite{msreview,yongmelt,yewc,clidensity}    of low-disorder 2DES at the low $\nu$ termination  of the fractional quantum Hall effect series.

         As a model of other stripe  
states in nature,  
the  striped phase of 2DES is of particular general interest, yet a complete microscopic picture of this phase is still lacking.   
The earliest  theory of the striped phases \cite{foglerkoulakov}, also incorporated a description of the bubble phase and IQHE-WC.  It described the the stripe phase as a unidirectional charge density wave, with charge uniformly distributed and liquidlike  along the direction of stripes of  width  a few classical cyclotron radii.    Other viewpoints of the striped phase were taken in subsequent theoretical work.   Descriptions \cite{fradkinlcstr,doanstratos} developed by analogy with smectic or nematic liquid crystals retain the  liquidlike   distribution of carriers along the stripes,    for the nematic over some finite distance.  
    In contrast, in   anisotropic Wigner crystal states \cite{fertigsmecstab,lifertigall,doanstratos,dorseyawc,maedastrcurrent},  also referred to as ``stripe crystals,''   the carriers are arranged in a  rectangular lattice and are much more closely spaced in the stripe direction, but still locked in position with respect to each other.

 Disorder pins  the stripe and isotropic electron solid phases,  so is of crucial importance to their phenomenology.   In the isotropic solids,  pinning operates in all directions, so that the electrons that form the 
solid are insulating.   In the stripe phase, pinning is expected at least in the direction perpendicular to the stripes,  resulting in $\rho_{xx}$, the resistivity
 in the hard direction, increasing as the temperature decreases.   Pinning along the stripes, as well,  is  expected for  the stripe crystal case \cite{fertigsmecstab,lifertigall,doanstratos,dorseyawc,maedastrcurrent}. 
Microwave spectra of isotropic electron solids 
\cite{yongmelt,yewc,clidensity,bubres,yongiqhwc,rupertcoex}, are known to be dominated by a  striking resonance that is understood as a pinning mode \cite{fertig,chitrawcall,foglerhuse},    in which pieces of  electron solid collectively oscillate about their pinned positions.     A pinning mode is induced by disorder, so  an increase in the disorder potential effectively experienced by the solid increases the frequency of the pinning mode. 
Pinning modes have long been known to exist in the archetypical Wigner solid  found at the low $\nu$ termination of the series of fractional quantum Hall effect states,    in low disorder n-type samples  \cite{yongmelt,yewc,clidensity}.  Similar pinning modes have been found in   the IQHE-WC \cite{yongiqhwc} and   bubble phases \cite{bubres,rupertcoex}.    

\begin{figure}[t]
   \includegraphics[width=2.8in]{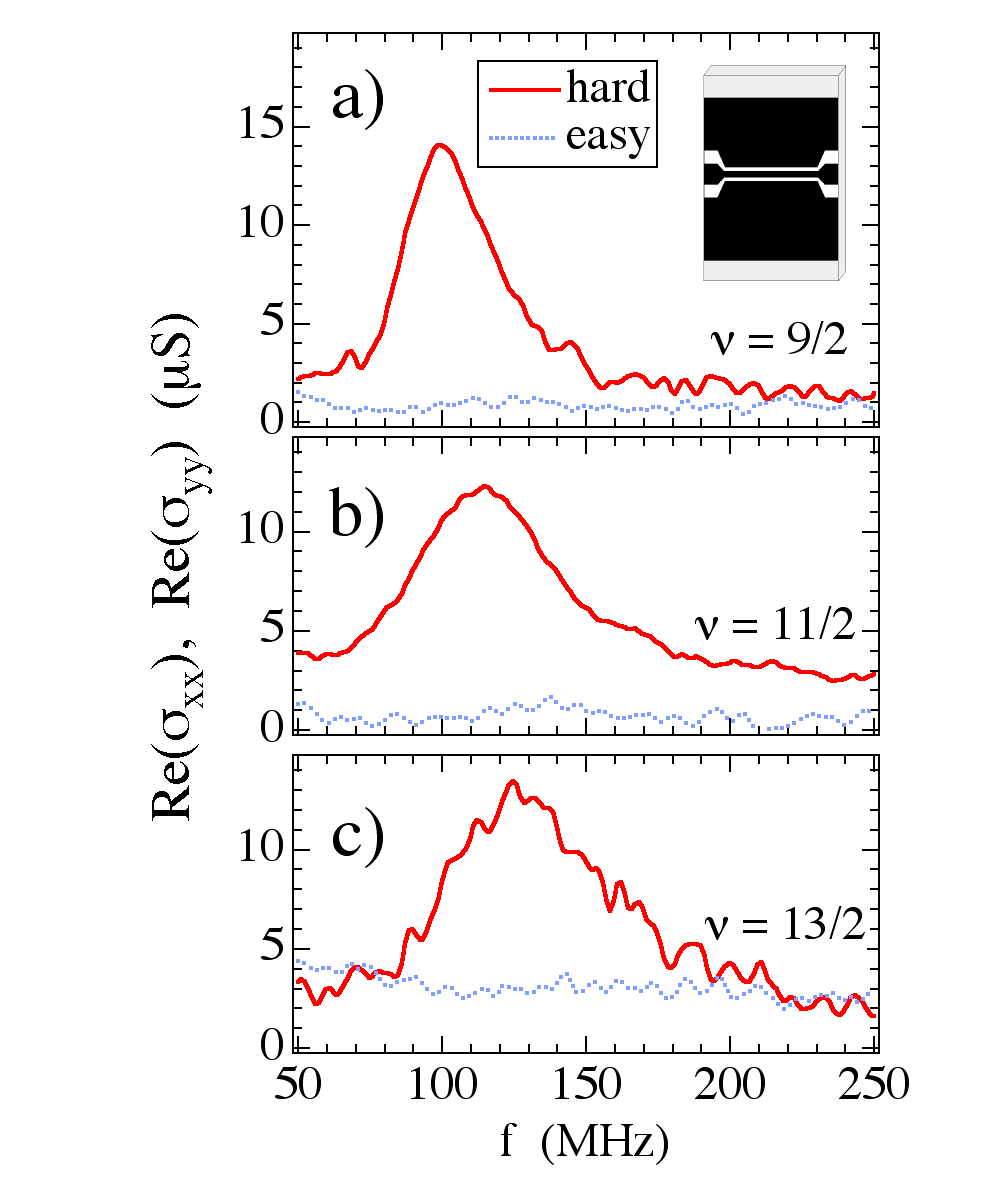}
\caption{   Spectra of real diagonal conductivities Re$(\sigma_{xx})$ (solid lines) and  Re$(\sigma_{yy})$ (dotted lines), vs frequency, $f$, for   filling factors  $\nu$ of a) $9/2$, b) $11/2$, and c) $13/2$.     Re$(\sigma_{xx})$  is measured in Sample 1, with the rf electric field along the hard direction, and   Re$(\sigma_{yy})$ is measured in Sample 2, with the rf electric field along the easy direction.   Inset in a) shows schematic of a sample,  with metal film that forms the  transmission line shown as black.    } 
\label{figallh}
\end{figure} 

In this paper we report  rf resonances in the spectra of   the stripe phases, around $\nu=9/2$ and higher half-integer fillings.  
  These resonances  are  discernable only when the rf electric field    is oriented along the higher dc resistance,  ``hard" direction, which  is  nominally perpendicular to the stripes.   These hard-direction   resonances  in the stripe phase    are most naturally understood as pinning modes, since   they     resemble   known, isotropic,  pinning modes in neighboring    bubble phase  regions.    Comparison is made to 
 theories \cite{lifertigall,orignac} of pinning modes in the stripe phases.    
      
As in earlier work 
\cite{yongmelt,yewc,clidensity,bubres,yongiqhwc,rupertcoex}, we obtained diagonal conductivities  from measurements of metal transmission lines lithographed onto the samples, and coupled  capacitively to the 2DES.  The transmission lines, of coplanar waveguide (CPW) type   have a narrow, driven centerline of lengths $l\sim 4$ mm, and grounded planes on either side separated by a slot of width $W=78\ \mu$m.  A  sketch of the CPW pattern   on a sample is inset  in  \fig{allh}a.  
The   transmission lines   
apply an rf electric field  {\em perpendicular} to their propagation direction.  
The CPWs are fixed on the sample surface, so we measured conductivities 
      $\sigma_{xx},\sigma_{yy}$ along orthogonal  host-lattice directions 
 $[1\bar{1}0]$ and $[1{1}0]$ ,
  using two samples made from adjacent pieces of the same wafer:   Sample 1 has rf electric field along the hard direction $[1\bar{1}0]$ and is used to measure   $\sigma_{xx}$, while sample 2   has rf electric field,  along  the  easy direction $[1{1}0]$ and is used to measure $\sigma_{yy}$.
      dc measurements on a third piece verified that $x$ and $y$ directions are respectively hard and easy.     The 2DES is in a 30 nm  GaAs quantum well,  and has  electron density, $n$ = 2.6 $\times$ 10$^{11}$  cm$^{-2}$ and mobility $\mu$= 2.9 $\times$ 10$^7$ cm$^2$/V-s   at   0.3 K.   The samples were not illuminated at low temperature, for either the rf  or the 
  dc measurements.  The data shown here were obtained at temperature $ \approx 35 $ mK,  and the rf power was varied to ensure the measurements were in the low-power limit.

\begin{figure}[t]
  \includegraphics[width=3.5in]{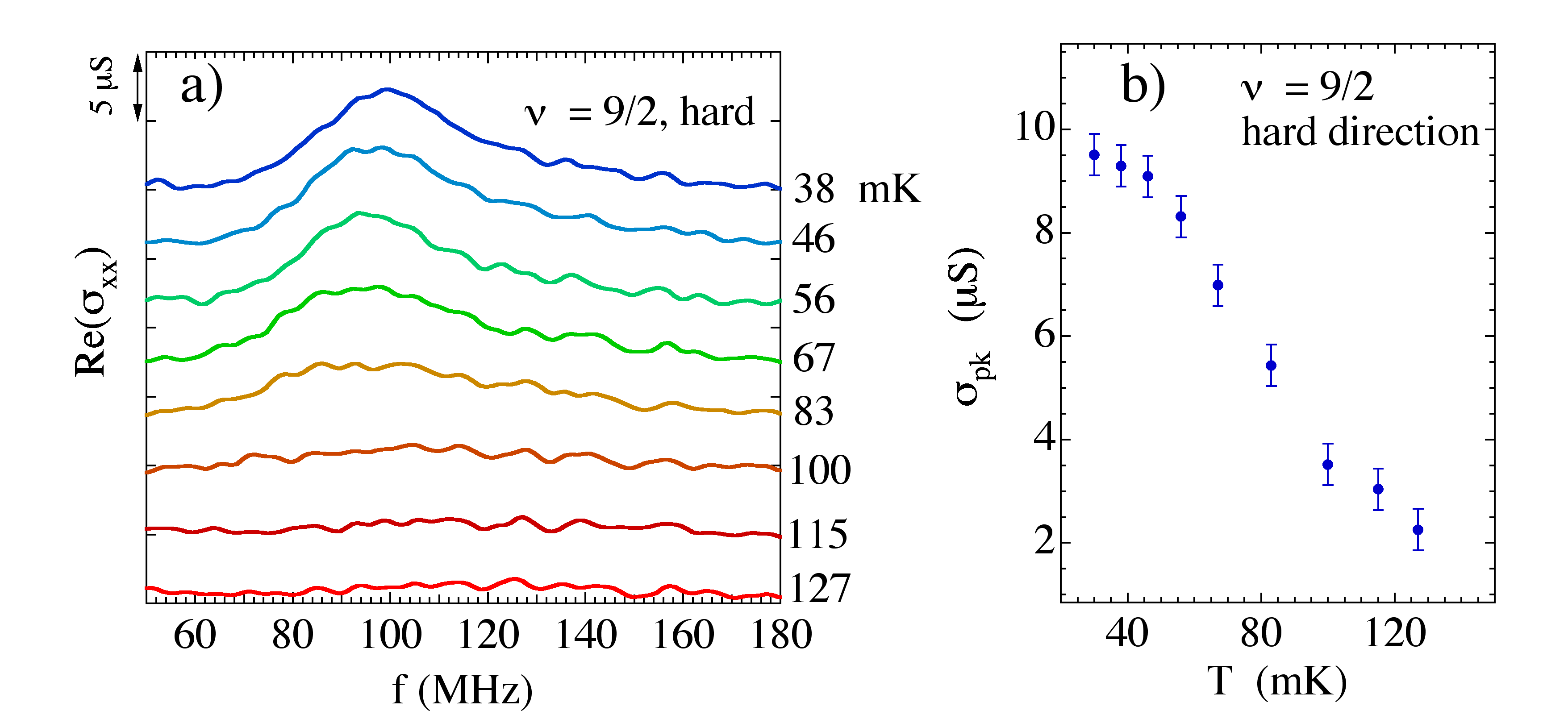}
\caption{ a)  Spectra of hard-direction real diagonal conductivities Re$(\sigma_{xx})$ at Landau filling $\nu=9/2$,  from sample 1, at  temperatures marked at right.   Successive spectra are offset upward by 4 $\mu$S for clarity.     b) Maximum Re$(\sigma_{xx})$, $\sigma_{pk}$, vs temperature $T$.   } 
\label{figtdep}
\end{figure}

We present real diagonal conductivities 
(Re$(\sigma_{jj})$ where $j=x$ or $y$)
  calculated from $t$, the complex measured amplitude transmission coefficient of the line,   normalized to conditions of vanishingly 
  small $| \sigma_{jj}|$.   
The plotted conductivities are calculated as $\sigma_{jj} =-W/(Z_0 l) \ln(t)$, where    $\ Z_{0}=50\ \Omega$ and is the characteristic impedance of the CPW calculated in the limit of vanishing $\sigma_{jj}$.  This formula takes the in-plane  rf electric field  to be confined in the regions under the  slots, which is  a good approximation for small enough  $\sigma_{jj}/f$.   A model employing  quasistatic calculation of the electric fields around the CPW in the presence of the conducting 2DES  was used to check the validity of the $\sigma_{jj}$ calculated in this way.     This model   does not allow for wave vector ($q$) dependence of  $\sigma_{jj}$, which  is thus taken to be in its low $q$ limit.  

 \fig{allh}   shows   the hard- and easy-direction real diagonal conductivity spectra, for  half integer fillings $\nu=9/2,\ 11/2  $ and $13/2$, at which the stripe phases are  known to exist.   The hard direction conductivity, Re$(\sigma_{xx}) $  is from sample 1, and the easy direction conductivity  Re$(\sigma_{yy})$ is from sample 2.     In each spectrum a clear resonance can be seen,   with the  peak frequency  $f_{pk}$ increasing slightly,  and the  resonance becoming broader and less well-developed at larger half integer $\nu$.  In the easy direction no resonance can be seen.    A weak resonance (not shown) was found in the hard direction at $\nu=15/2$ as well.

 \fig{tdep} shows the $\nu=9/2$  resonance in  the hard direction at various temperatures. 
  The resonance in  the hard direction    disappears gradually as temperature increases, becoming invisible   above $\sim  120 $ mK.  This is about the same temperature range that was observed for the  dc resistance anisotropy \cite{lillyaniso,duaniso}, consistent with the resonance being due to formation of the stripe phase.    As is typical of pinning modes in the low $\nu$ Wigner solid range \cite{yongmelt,clitdep},  $f_{pk}$ is nearly independent of the   temperature. 
 
\fig{carpet}   shows  series of spectra with $\nu$ ranging from   $4.38$ to $ 4.65$; the spectra are  vertically offset  proportional  to  $\nu$.  \fig{carpet}a
shows the hard-direction conductivity, Re$(\sigma_{xx})$, from sample 1, while  \fig{carpet}b
shows easy-direction conductivity,  Re$(\sigma_{yy})$, from   sample 2. 
For $\nu$ from  $4.41$ through   $4.60$,  a resonance is present in the hard direction, while none can be discerned in the easy direction. 
 This  range  is in reasonable  agreement with theoretical predictions 
\cite{shibata,goerbigcompete,cotebubdyn,dorseyawc}  for the occurrence of the stripe phase in the $N=2$ LL;  the predicted  low $\nu$ crossover between stripe and bubble phase ground states falls between     4.35  \cite{goerbigcompete} to 4.43   \cite{cotebubdyn}.   As $\nu$ decreases from 4.41, or increases from 4.60, the   resonance in the easy direction develops rapidly, and by $\nu=4.37$ and $ 4.63$,  the highest and lowest  $\nu$ for which spectra are shown,  the spectra are nearly isotropic,  so we regard these fillings as in the bubble phase.

    \begin{figure}[t]
  \includegraphics[width=3.5in]{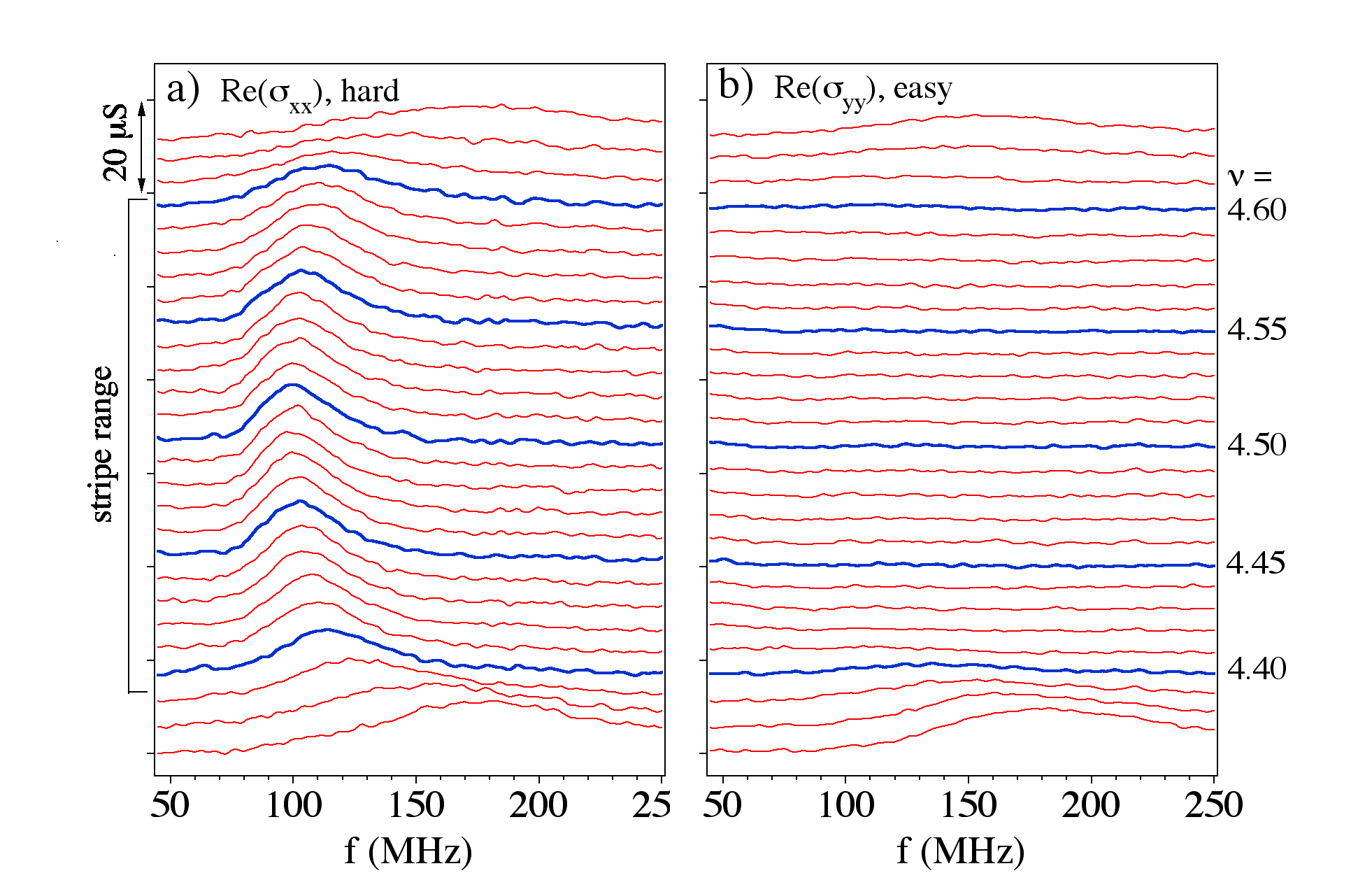}
\caption{     Spectra, conductivity  vs frequency, $f$, for many filling factors $\nu$ between 4.37 and 4.63; each succeeding trace  offset upward by 5 $\mu$S:  a)  Hard direction conductivity   Re$(\sigma_{xx})$ measured in sample 1;    b) Easy direction conductivity, Re$(\sigma_{yy})$, from sample 2.  
 Filling factors $\nu$  are marked at right.     The stripe range predicted in ref \cite{shibata} is marked at left. } 
\label{figcarpet}
\end{figure}

\fig{sum} shows  plots vs $\nu$ of the resonance frequency  ($f_{pk}$),  the  peak   real diagonal conductivity ($\sigma_{pk}$), and the resonance quality factor ($Q$,   $f_{pk}$ divided by full width at half-maximum).
  We estimate errors in $\nu$ 
 at $\pm 0.02$ and errors in $\sigma_{pk}$ as $\pm 0.5 \ \mu$S; errors in $f_{pk}$ and $Q$ are as shown.   At $\nu= 9/2$, the hard-direction $f_{pk}$  vs $\nu$  exhibits a shallow minimum, while $Q$  exhibits a weak maximum.   
 Within most of the stripe range, from $\nu=4.41$ to $4.59$,  $\sigma_{pk}$, is flat within  the experimental error.  
For $\nu\le 4.40$ or $\nu\ge 4.60$, the    easy-direction resonance is observable:  its $f_{pk}$ agrees with that of the  hard-direction resonance to within experimental error,  though its $\sigma_{pk}$ decreases rapidly as $\nu$ moves toward the center of the plot.

The natural interpretation of the    resonance in the stripe phase  is as a pinning mode. The stripe phase resonance appears to evolve out of known pinning mode resonances present in the neighboring bubble phases, which themselves succeed resonances in the IQHE-WC, as described in ref. \cite{rupertcoex}.     
A resonance is present in the hard direction throughout the range from $\nu=4.10$ to $4.90$.      At $\nu=4.10$ the  resonance is well-understood as a pinning mode of the IQHE-WC. Increasing $\nu$ results in a transition to a bubble phase, clearly visible as    a lower $f_{pk}$ resonance \cite{rupertcoex}.  As apparent in \fig{sum}a,  $f_{pk}$  drops again at the transition from bubble phase to stripe  phase.      $\sigma_{pk}$  changes less than 30 \% between the isotropic, bubble-phase  resonance at $\nu=4.63$  to the anisotropic stripe phase resonance  at $9/2$, strongly suggesting a similar    origin of the resonances.

  \begin{figure}[t]
  \includegraphics[width=2.2in]{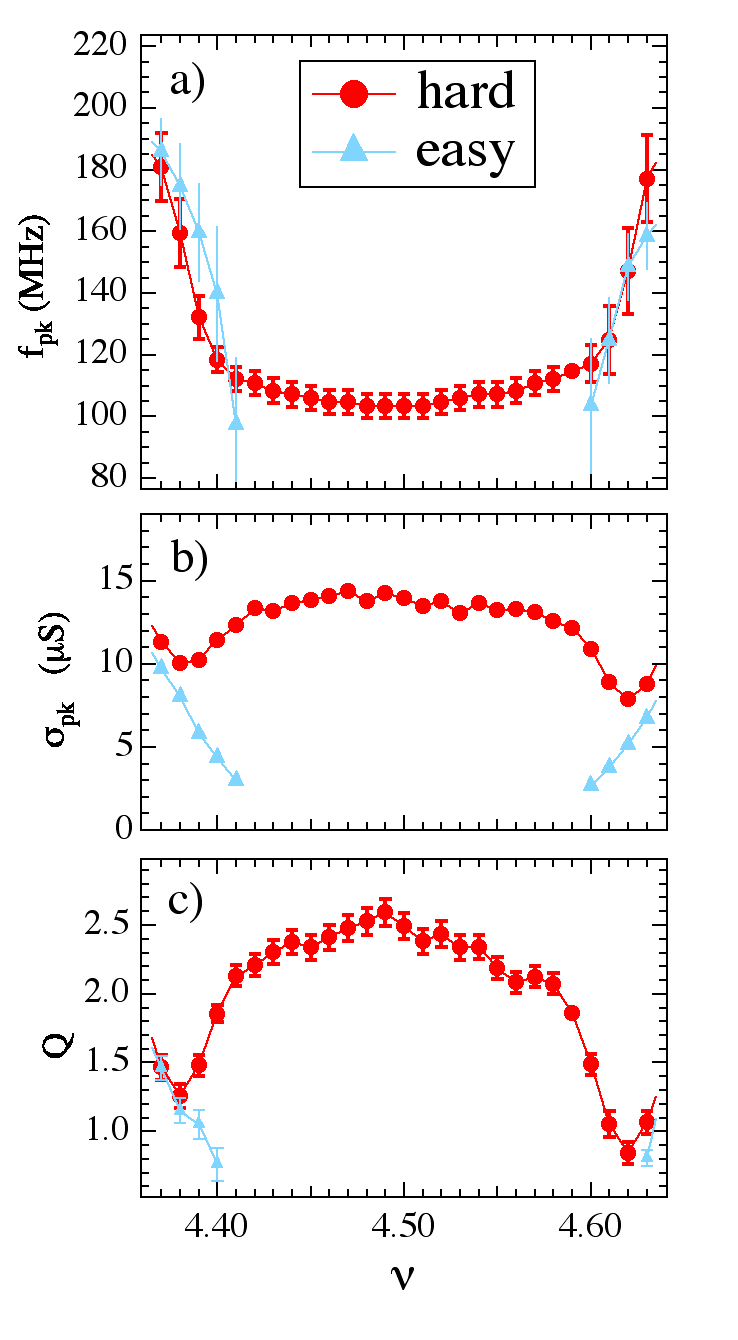}
\caption{  Peak frequency ($f_{pk}$, maximum real diagonal conductivity ($\sigma_{pk}$) and   quality factor $Q$ for the resonances,   vs $\nu$, for the hard and easy directions,    around $\nu=9/2$.   \label{figsum}}
\end{figure}

  A number of theories \cite{aoyama,dorseyawc,cotefertigstrmodes,lopatstr,fradkinnemmode,foglervinokur} treat modes of the stripe phase in the absence of disorder.  
If the observed resonance were not a pinning mode,  so that disorder did not play a role,  
$f_{pk}$ would be determined by the dispersion of the mode and a fundamental wavelength $W/2$ set by the transmission line  slot width, $W=78\ \mu$m.
 Particularly   in  treatments \cite{lopatstr,foglervinokur} based on  the  smectic liquid crystal,    propagating modes are present only at oblique angle to the stripes, not  perpendicular or parallel to them; this would be inconsistent with our finding of the mode in the hard direction.

  There have been to our knowledge only two calculations of pinning modes of the stripe phase with disorder \cite{orignac,lifertigall}.  Though both    \cite{orignac,lifertigall}  predict  pinning modes for rf electric field parallel to the stripes,   in   contrast with our not observing an easy-direction resonance, the theories indicate that easy-direction pinning mode can  be 
comparatively weak under our experimental conditions.    In 
ref.  \cite{lifertigall},  pinning modes are found only in a state  \cite{fertigpricomm} that has pinning both along the stripes and perpendicular to them.   The predicted pinning has about the same frequency  parallel  and perpendicular to the stripes, though the mode parallel to the stripes  is weaker by a factor of about 4.      Ref.  \cite{orignac} obtains only   the conductivity  along the stripes, and predicts the easy-direction pinning  mode frequency and amplitude both vanish at  small wave vector $q$.

In weak pinning of an electron solid, the pinning mode frequency  
$f_{pk}$ is larger either for increased  disorder or  reduced  solid stiffness,   
\cite{chitrawcall,foglerhuse,fertig}.  Such reduction in solid stiffness  can be realized in the low $\nu$, lowest LL Wigner crystal, by reducing the overall sample carrier density \cite{clidensity} and the increase in $ f_{pk}$ is accompanied by a decrease in $Q$.   ($f_{pk}$ is larger for a 
softer solid, since the carriers effectively fall more deeply into the disorder potential, increasing the average pinning.)      In this framework, the shallow minimum of $f_{pk}$ vs $\nu$ at 9/2,  as well as the maximum  in $Q$   is   due either to 
 a maximum of the  stiffness of the stripe phase at that $\nu$,  or to a minimum in the effective disorder strength.

 In summary, we have found an rf resonance that is clearly associated with the stripe phase of higher LLs, and occurs only for electric field along the hard direction.  The resonance appears to evolve from known pinning modes in the bubble phases, and is similar to them,  and so is naturally interpreted   as a pinning mode of  the stripe phase.

We thank  R. C\^{o}t\'{e}, A. Dorsey, and  H. Fertig  for discussions. This work was supported by  DOE Grant Nos. DE-FG21-98-ER45683 at Princeton, DE-FG02-05-ER46212 at NHMFL.  NHMFL  is supported by NSF Cooperative Agreement No. DMR-0084173, the State of Florida and the DOE.   

 \bibliography{strbib.bib}

\begin{thebibliography}{31}
\bibitem[{fer()}]{fertigpricomm}
\bibinfo{note}{Ref. \protect\cite{lifertigall} describes a transition within
  the stripe phase as $\nu$ approaches the half integer, from the state pinned
  in both directions to a state that is depinned along the stripes. Per private
  communication from H. Fertig, for larger disorder than that taken in that
  reference, the pinned state may well persist throughout the stripe range,
  including, e.g., $\nu=9/2$.}
\end{thebibliography}

\end{document}